﻿

\documentclass{article}
\usepackage{epsfig}     
\newcommand{\ltsimeq}{\raisebox{-0.6ex}{$\,\stackrel 
        {\raisebox{-.2ex}{$\textstyle <$}}{\sim}\,$}} 
\newcommand{\gtsimeq}{\raisebox{-0.6ex}{$\,\stackrel
{\raisebox{-.2ex}{$\textstyle >$}}{\sim}\,$}} 

\begin{document}

\begin{centering}
\Large\bf The hazard from fragmenting comets
\vspace*{0.35in}

\large \bf W.M. Napier\footnote{bill.napier@armagh.ac.uk}
\vspace*{0.25in}

\small{Armagh Observatory and Planetarium, College Hill, Armagh BT67 9DG, Northern Ireland, UK}
\end{centering}
\bigskip

{\small Accepted 2019 June 24. Received 2019 June 16; in original form May 26}

\begin{abstract}
Comet disintegration proceeds both through sublimation and discrete splitting events. The cross-sectional area of material ejected by a comet may, within days, become many times greater than that of the Earth, making encounters with such debris much more likely than collisions with the nucleus itself. The hierarchic fragmentation and sublimation of a large comet in a short period orbit may yield many hundreds of such short-lived clusters. We model this evolution with a view to assessing the probability of an encounter which might have significant terrestrial effects, through atmospheric dusting or multiple impacts. Such an encounter may have contributed to the large animal extinctions and sudden climatic cooling of 12,900 years ago, and the near-simultaneous collapse of civilisations around 2350 BC.
\end{abstract}
\smallskip

\bf{Keywords:} comets -- Earth -- interplanetary medium

\section{Introduction}
It has been proposed that the onset of the sudden cooling at the Younger Dryas boundary (YDB) of 12,800$\pm$150 years ago was due to a celestial encounter. This is supported by the presence of high concentrations, at the boundary, of platinum-rich dust at thirty sites throughout the northern hemisphere (Petaev et al. 2013; Moore et al. 2017), along with a wide northern hemisphere distribution of claimed impact proxies such as glassy microspherules (Bunch et al. 2012), nanodiamonds (Bement et al. 2014) and an estimated ${\sim}10^7$ tons of magnetic spherules argued to be of impact origin (Wittke et al. 2013). These proxies are often found in combination with major changes in fauna and flora indicative of sudden climate cooling (Kletetschka et al. 2018).

Evidence has also been presented that the largest biomass-burning event of the last 120,000 years, with perhaps 5-10\%  of the Earth's biomass being consumed in wildfires over a few days or weeks, occurred at the YDB (Wolbach et al. 2018). Many large animal species became extinct in the same geological instant; in North America up to 35 large animal genera in North America were extinguished almost simultaneously (Faith \& Surovell 2009); in Australia, the extinctions amounted to about 80\% of large species. The cause of these megafaunal extinctions, and their precise timing, have been disputed (Scott et al. 2017), as has the interpretation of the ground evidence in terms of a cosmic encounter: these arguments and references are summarised by Pino et al (2019). The northern hemisphere cooling took hold within a few years or less, the temperature drop being comparable with a return to glacial conditions, and persisted for $\sim$1300~yr. 

Pino et al (2019) have extended the investigation of the YDB to a southern hemisphere site in Chile, 40$^\circ$S, and again find a major peak in charcoal abundance, with evidence of megafaunal extinctions synchronous with those in the northern hemisphere, and similar elemental peaks such as platinum, gold and high-temperature iron spherules, taken to be cosmic input proxies. The cosmic input, if real, thus seems to extend over at least hemispheric dimensions.

With currently adopted impact rates, there is an expectation of one near-Earth asteroid impact of energy ${\sim}200$~megatons over the last 20,000~yr, quite inadequate to produce the observed global trauma  (Bland \& Artemieva 2006). The impact of a 4~km comet so recently in the past, as had originally been proposed for the event (Firestone et al 2007), is \textit{a fortiori} improbable. A subglacial impact crater ${\sim}30$~km in diameter in north-west Greenland has been attributed to the impact of a 1.5~km iron asteroid some time during the Pleistocene (Kjaer et al. 2018). Its precise age is, however, unknown and so its relation to the Younger Dryas geology is at present uncertain.

There are several lines of evidence to indicate that a large (${\sim}100$~km) progenitor comet in a low inclination, short period orbit was at that time giving rise to Comet 2P/Encke and the Taurid meteors through a cascade of disintegrations (Clube \& Napier  1984).  Backtracking the meteor orbits, Steel \& Asher (1996) estimate the initial disintegration to have begun ${\sim}20,000$ years BP  --  although an earlier epoch cannot be discounted --  and it has been suggested that the Younger Dryas boundary phenomena might have been triggered by an encounter with some of the fragmented material (Napier 2010). The object of this paper is to model the disintegration of the progenitor comet in more detail, to see whether a plausible match can be made between the astronomical environment of that time and the terrestrial record; and hence, more generally, to discuss the role of such comets in past climate variations.

\section{Comet disintegration modes}
Di Sisto et al. (2009) showed that there is a poor match between the distribution of Jupiter family orbital elements expected dynamically, and that observed, and attributed the difference to the disintegration of comets through sublimation and splitting. They derived formulae for both processes and gave four `good match' models (Table\,1) capable of reconciling the dynamics and the observations.  These are statistical models for an ensemble of Jupiter family comets, but are taken below to apply to the progenitor of the Taurid meteor complex and Comet Encke. The fraction of mass lost in a splitting event is, according to di Sisto et al., given by 
        \begin{equation}
        s = s_0/R
        \end{equation}
        \noindent $R$ the comet radius in units of 10~km.
        \medskip
        
        The probability of splitting per revolution is 
        \begin{equation}
        f = f_0 (q/q_0)^{-\beta} 
        \end{equation}
        \noindent with $q_0 = 0.5$\,au. 
        \medskip

\begin{table}

\begin{tabular}{rrrr}
${\rm model}$         & $\beta$   &  $f_0$  &  $s_0$   \\ \hline
1                               &  1             &   1/4     &  0.01      \\
2                               &  1             &   1/3     & 0.007     \\
3                               &   0.5         &   1/6     & 0.01       \\
4                   &   0.5         &     1       & 0.001      \\ \hline
\end{tabular}

\caption{\small Comet fragmentation models from Di Sisto et al (2009).  The observations can be fitted by frequent splitting events $f$ with small mass loss $s$ per event (model 4), or fewer events with larger mass loss (model 1).}
\end{table}     

\medskip

\begin{table}

\begin{tabular}{rrrr}
${\rm model}$ & final radius &   $N_f$     & $T$ \\ \hline
1                       &  16  &   670  &    5900         \\
2                       &  19  &   830  &  10000         \\
3                       &  18  &   650  &  20000         \\
4           &  16  &       0  &    ---      \\ \hline
\end{tabular}

\caption{Simulations of comet destruction, with sublimation and random splittings. The comets are in an Encke-like orbit with initial diameter 100~km and mass $4.7e20$~gm. The number of splittings yielding fragments of total mass $1e17$~gm  is recorded ($N_f$),  along with the time taken to evolve to the state where the fragment swarms have mass $\le 1e 17$~gm.}
\end{table}     

\medskip

\begin{figure}
\noindent
\begin{minipage}[b]{\linewidth}
\includegraphics[angle=270,width=\linewidth]{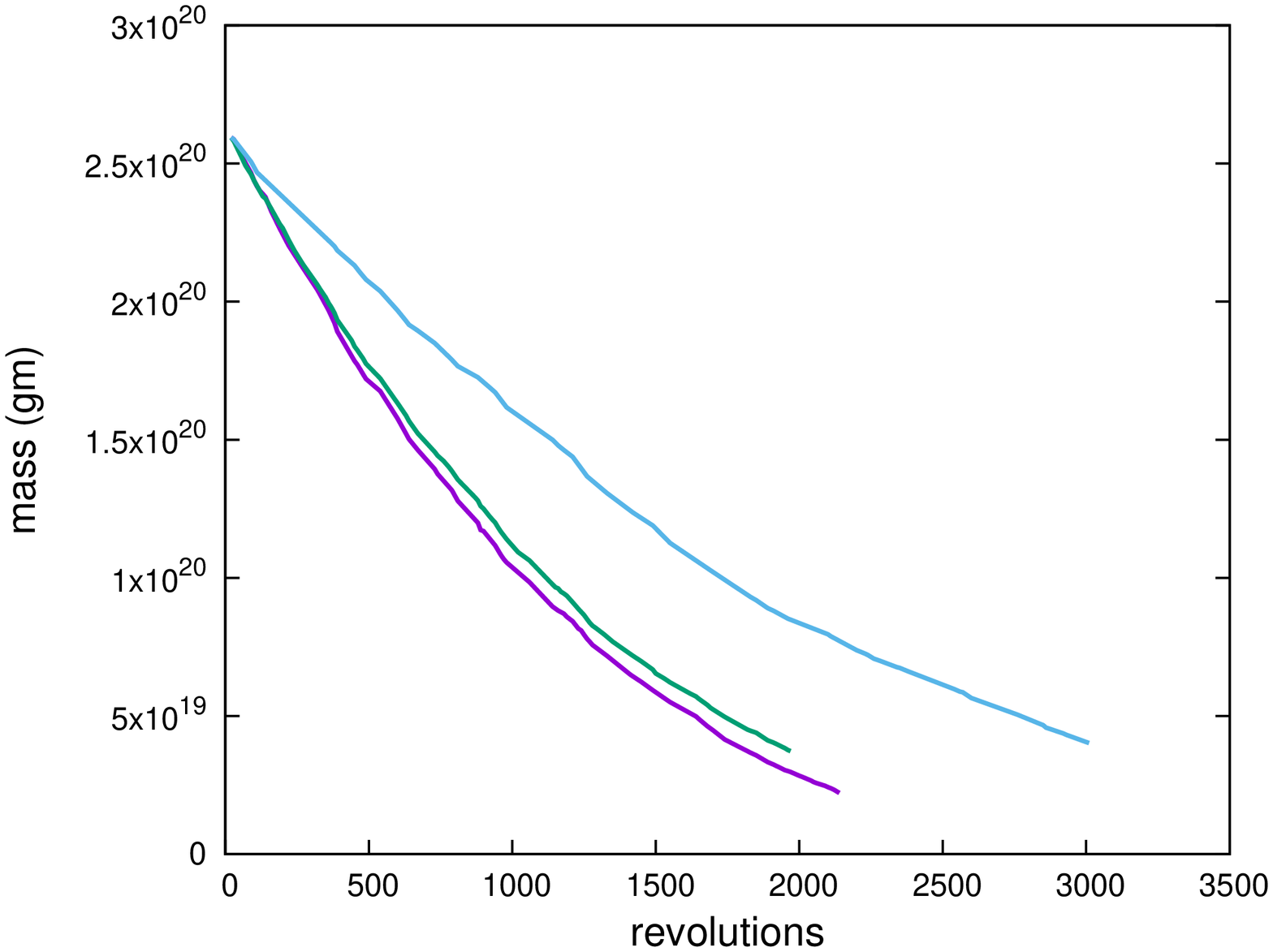}
\end{minipage}
\noindent
\begin{minipage}[b]{\linewidth}
\includegraphics[angle=270,width=\linewidth]{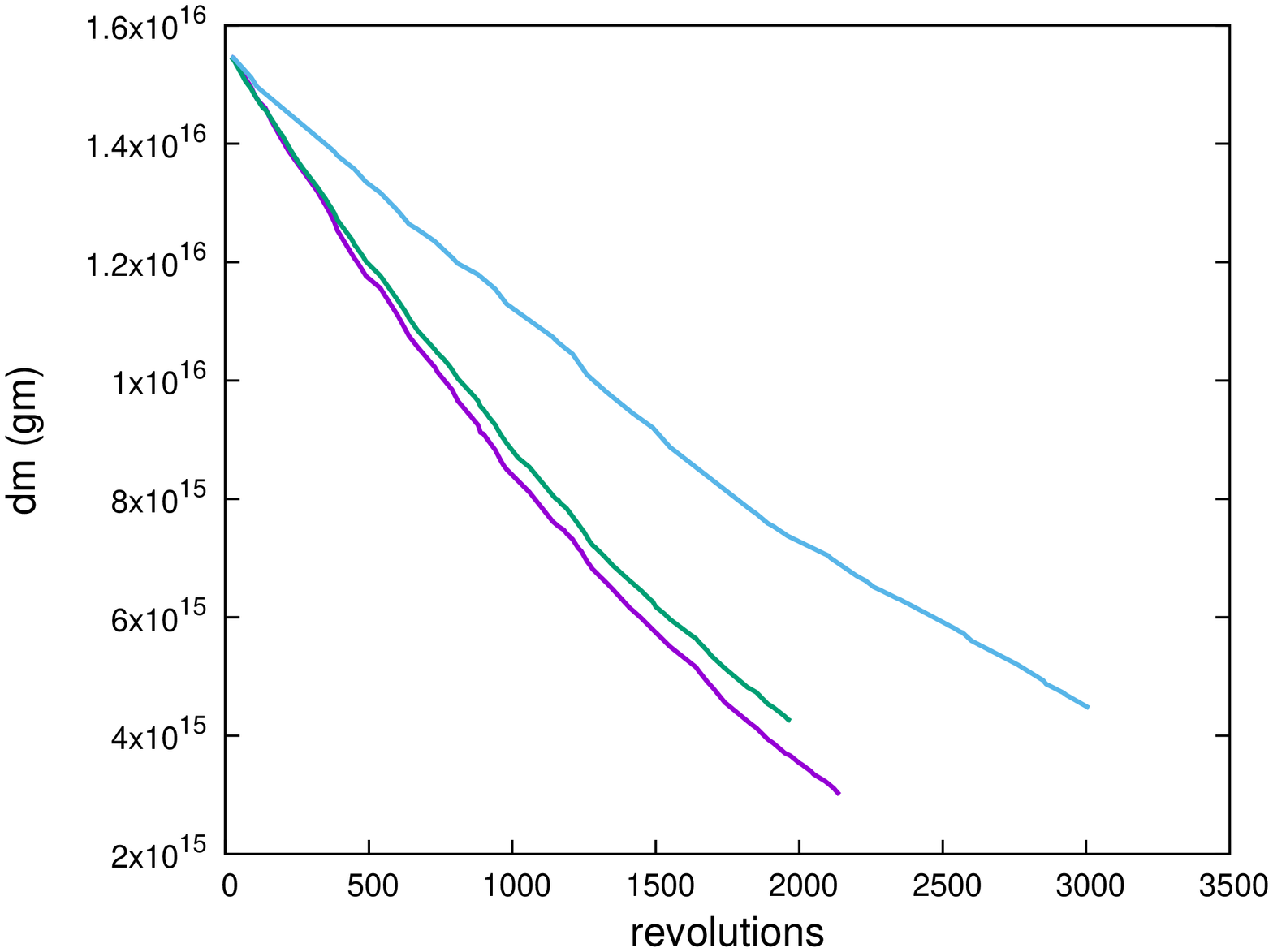}
\end{minipage}
\caption{Mass evolution of a 100 km comet in an Encke-like orbit. Models 3,1,2 top to bottom. Evolution is followed for 10,000 years. Mass lost by sublimation (bottom graph) is four orders of magnitude less than that lost by splitting (top graph).}

\end{figure}

\begin{figure}
   \includegraphics[angle=270,width=\linewidth]{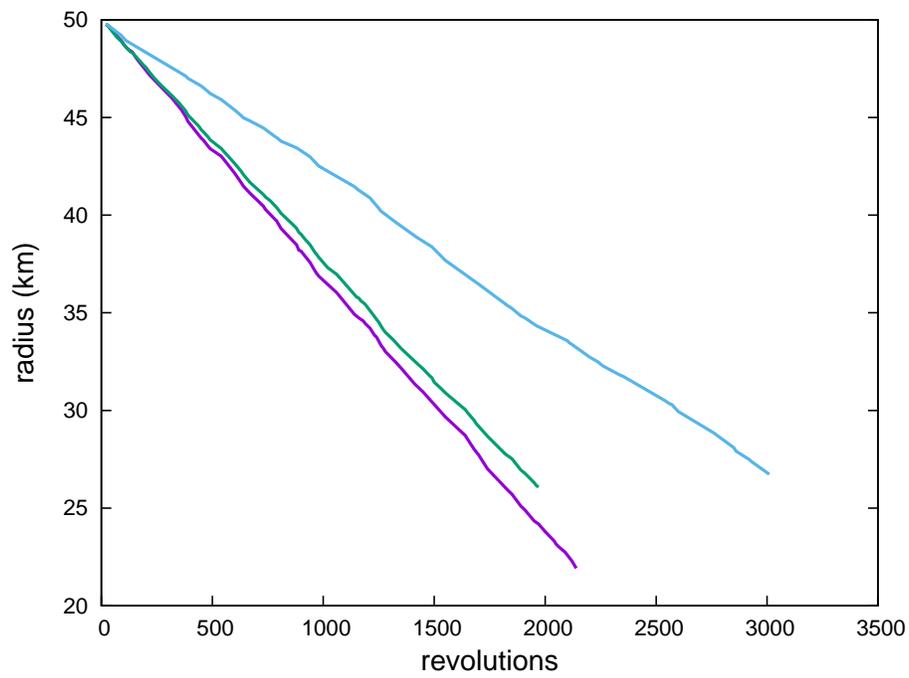}
\caption{Radius evolution for models 1--3.}
\end{figure}

\begin{figure}
   \includegraphics[angle=270,width=\linewidth]{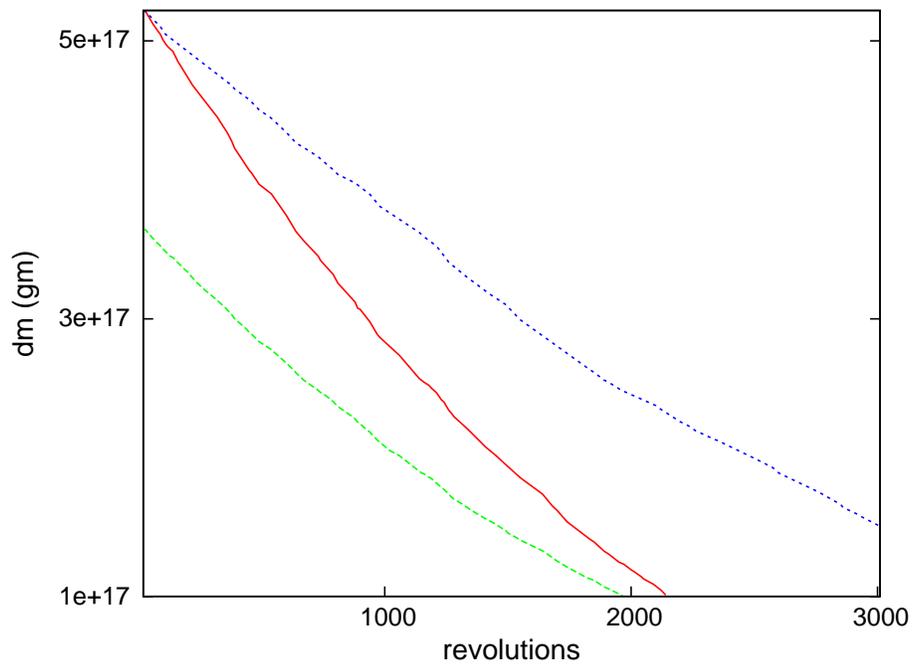}
\caption{Mass of material ejected at each splitting event, following the evolution of the 100 km comet.}
\end{figure}
        
Figures 1 and  2 show, respectively, the evolution of mass and radius of a 100~km comet in an Encke-like orbit ($q=0.34$~au), simulated by way of the parameters given in models 1 to 3.  The splittings took place randomly, in accord with the probabilities given by eqn (2), over a period of 10,000 years.  Splittings with mass loss ${\ge}10^{17}$\,gm took place 779, 939 and 611 times for models one, two and three respectively. These occur anywhere along its orbit but may tend to concentrate around perihelion (Boehnhardt 2004). Splitting is seen to be the dominant process of comet disintegration, with mass loss by sublimation four orders of magnitude down (Fig.~1). As comet disintegration proceeds and its mass dwindles,  the mass of fragments split off likewise declines (Fig.~3). Model 4, with very frequent low mass splittings, produced no fragment swarms with mass $>10^{17}$\,gm.

\section{Ejection and spread of fragments}
\begin{table}
\centering

\begin{tabular}{rrrrr}
${\rm model}$ & $D_0$ & $L_{rev}$ &  $L_{yr}$ &   $N_f$     \\ \hline
1                       &    150  &  5560     &   18350     &    1474      \\
1                       &    100  &  3700     &   12230     &     776      \\
1                       &     50  &  1850      &    6090     &      84      \\
1                          &     20  &   700        &    2300     &       0       \\  \hline
2                       &    150  &  6050     &   20000     & 1924     \\
2                       &    100  &  4030     &   13285     &  943      \\
2                       &     50  &  2025      &    6675     &    0     \\
2                       &     20  &   740        &    2450     &    0     \\ \hline
3                       &    150  & 6050      &   20000     & 1257     \\
3                       &    100  & 6050      &   20000     &  734      \\
3                       &     50  & 6050       &   20000     &   83     \\
3                       &     20  & 1160       &    3830     &    0     \\ \hline
\end{tabular} 
\caption{The physical evolution of a comet in an Encke-like orbit with initial diameter $D_0$ km. The number of splittings yielding fragment clusters of mass $>10^{17}$~gm  is recorded ($N_f$),  along with the lifetime of the progenitor. These are measured in revolutions $L_{rev}$ or years respectively $L_{yr}$ and are taken to last until the diameter of the comet declines to $<$1~km.}
\end{table}     

\medskip        

I carried out simulations in which comets of various sizes in an Encke-like orbit underwent sublimation and random splittings in accord with the di Sisto et al. models, following the evolution for 20,000 years. Table~2 shows the lifetimes and numbers of splittings with masses in excess of $10^{17}$~gm.  The lifetime of a comet was measured by the time taken to decline to a $<$1~km object. All the models, except 4, predict something like 750-1500 splitting events yielding fragment clusters each of mass $>10^{17}$~gm, over timescales of order 6000 - 20,000 years (Table~3).  At an encounter speed of  $\rm {\sim}30\,km\, s^{-1}$, 10$^{17}$~gm of debris carries $10^7$ megatons of impact energy.

\section{Encounters with fragment swarms}
For the sake of illustration assume that the orbital elements adopted for the Taurid progenitor are those for the current Comet Encke: 
$a$ = 2.214 au, $e$ = 0.84834, $i$ = $11.783^\circ$. From
\begin{equation}
r = \frac{a(1-e^2)}{1+e\cos \phi} 
\end{equation}

\noindent With $r=1$~au, encounters with the Earth are possible when the true anomaly $\phi = 116.56^\circ$ or $243.44^\circ$. These intersections occur $(42.3+ nP)$ and $(1160.9+nP)$ days after perihelion, on the outward and inbound legs of the orbit respectively, $n$ the number of complete revolutions and $P{\sim}$1203 days the orbital period. The encounter speed is $\rm {\sim}30~km\, s^{-1}$. The probability of an encounter, and the energetics, are functions of the time of splitting, the ejection speed of the fragments, their total mass, and the epoch of the encounter.

\begin{figure}
   \includegraphics[angle=0,width=\linewidth]{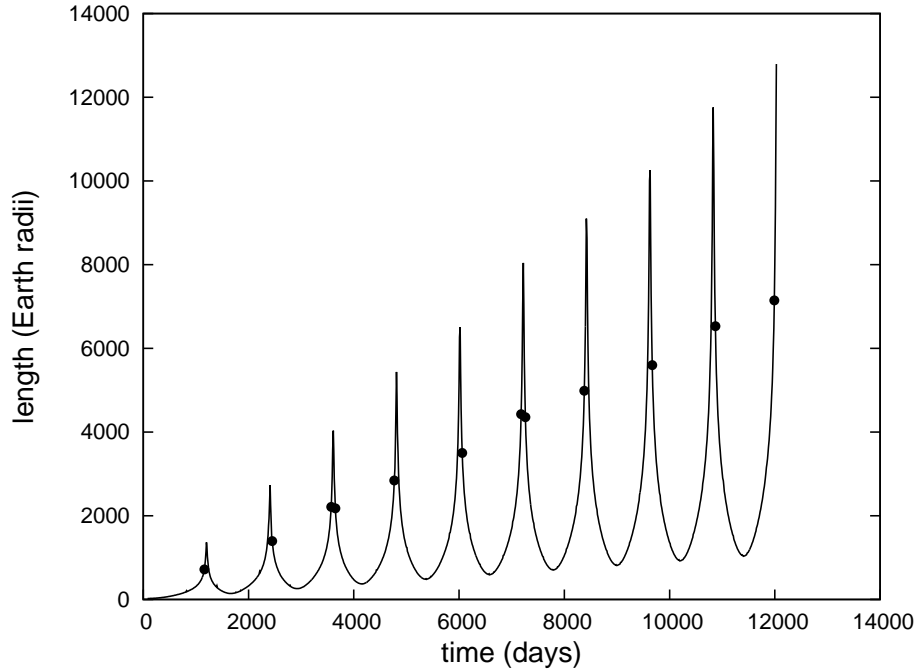}
\caption{Evolution of trail length over ten orbital periods (33~yr), for a $\rm 2\,m\,s^{-1}$ maximum dispersion of fragments. The peak lengths occur at perihelion passages, superimposed on a secular increase in trail length. Passages at 1~au are marked by dots. Inbound and outbound adjacent trail lengths are approximately equal.}
\end{figure}

\begin{figure}
   \includegraphics[angle=0,width=\linewidth]{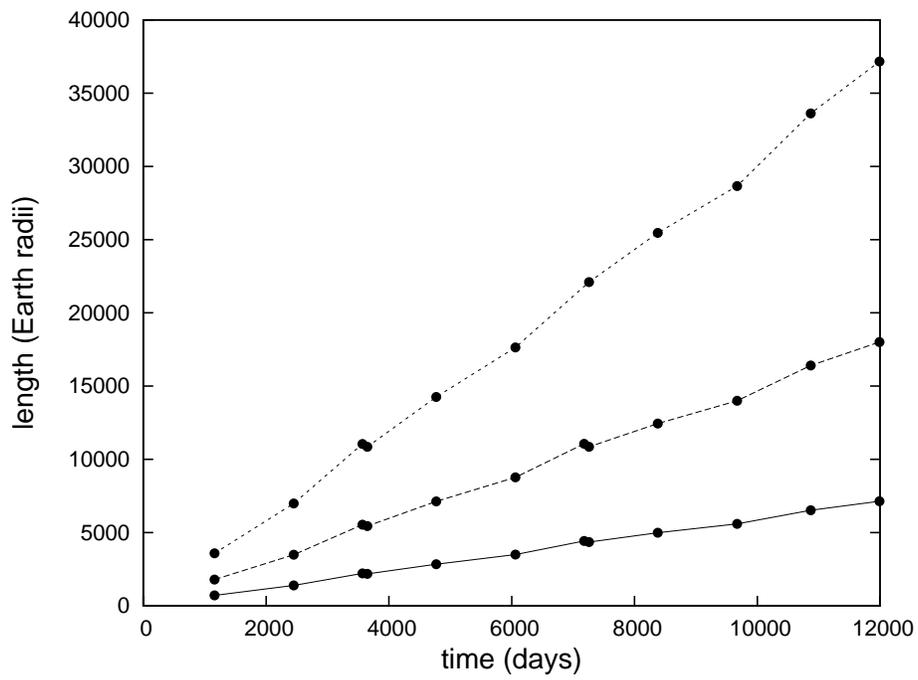}
\caption{Evolution of trail lengths over ten orbital periods (33~yr), for maximum dispersion speeds 10, 5 and 2 $\rm m\,s^{-1}$ (top to bottom). Passages at 1~au are again marked by dots.}
\end{figure}

\begin{figure}
   \includegraphics[angle=0,width=\linewidth]{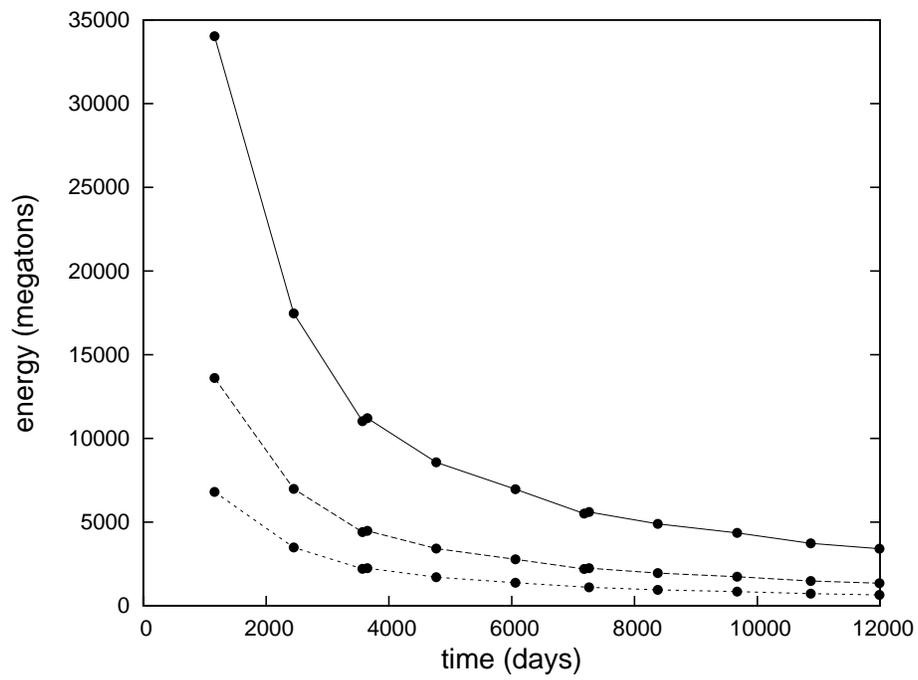}
\caption{Kinetic energy of encounters with a debris trail of mass $10^{17}$\,gm for 10 comet revolutions (${\sim}33$~yr). The trail disperses at 2, 5 and 10 $\rm \,m\,s^{-1}$ respectively, top to bottom.}
\end{figure}

Simulations were carried out in which the fragments from a breakup were dispersed isotropically with random dispersion speeds up to a maximum $\Delta v\rm \,m\,s^{-1}$.  Estimates of the ejection velocities of meteoroids from comets are generally in the range $\rm 0-50 \,m\,s^{-1}$, although IRAS observations of cometary dust trails yield values less than $\rm \pm 5 \,m\,s^{-1}$ (Sykes \& Walker 1992, Kres\'{a}k 1993). Boehnhardt (2004) finds that, for short period comets, a mean of 16 cometary fragments is generated during a spitting event, with separation velocities  $\rm 2.7\pm2.3\,m\,s^{-1}$. Fig.~4 illustrates the evolution of length of a debris trail over ten orbits. This was formed by a splitting at $\phi=135^\circ$, 80 days after perihelion passage t, with $\rm \Delta v =  2\,m\,s^{-1}$ debris. The trail stretches and shrinks as it passes through perihelion and aphelion respectively, but superimposed on this is a secular lengthening at a rate of $\rm {\sim} 43\,m\,s^{-1}$. The rate of lengthening of the trail is directly proportional to the initial breakup speed of the fragments, but was found to be not strongly dependent on the true anomaly at which fragmentation occurred, apart from those taking place within a few days of perihelion. Fig.~5 illustrates the evolution of trail lengths at 1~au for several dispersion speeds.  The trail is 1\,au from the Sun 42.3 days before or after its perihelion passages.  

 Table~3 reveals that the number of trails created is in the hundreds over the physical lifetime of the comet. As material in a trail disperses along the comet orbit, the probability of a terrestrial encounter with it increases but its incident kinetic energy, in the event of an encounter, declines. This is illustrated in Table~4, which gives encounter frequencies and energies for a swarm of fragments with mass $10^{17}$~g and maximum dispersion velocity $\rm \Delta v = 2 \,m\,s^{-1}$ as it spreads along the orbit. $L$ is the length of the swarm, which travels through the node at 1~au at a speed $\rm V{\sim}23~ km\,s^{-1}$. For the dispersions considered, the vertical dispersion of the material is less than $R_\oplus$ and the mass $M_{enc}$ intercepted by the Earth is $2R_\oplus/L/\cos\chi$, $\chi{\sim}64.4^\circ$ the angle at which the Earth intercepts the swarm.  The first 15 passages at 1~au would yield, in the event of an encounter with Earth, a mass influx with kinetic energy in excess of 3000 megatons (1~Mt${\sim} 4.2 \times 10^{22}$~ergs). Fig.~6 illustrates the impact energies for the first ten potential encounters following splitting events in which the fragments initially disperse at  2, 5 and 10~$\rm m\,s^{-1}$. 
\begin{table}
\centering
\begin{tabular}{rrrrrr}
    yr &   $L$   & $D$ &  $\nu$ & $M_e$ &   $E$   \\ \hline
   3.3 &   732   & 2.4  & 0.002  & 40.3   & 42.8  \\
    6.6 &   1372 & 4.4  & 0.004 & 21.5   & 22.8  \\
   9.9 &   2012 & 6.5  &  0.005 & 14.7   & 15.6  \\
   13.2 &   2652 & 8.5  & 0.007 & 11.1   & 11.9  \\
   16.5 &   3292 &10.6 & 0.009 & 9.0   & 9.1  \\
   19.8 &   3932 &12.7 & 0.011 & 7.5   & 8.0  \\
   23.1 &   4572 &14.7 &  0.012 & 6.5   & 6.9  \\
    26.4 &   5212 &16.8 & 0.014 & 5.7   & 6.0  \\
   29.6 &   5852 &18.8 &  0.016 & 5.0  & 5.8  \\
  32.9 &   6492 &20.9 &  0.017 & 4.5   &  4.8  \\
  36.2 &   7132 &23.0 &  0.019 & 4.1   &  4.4  \\
  39.5 &   7772 &25.0 &  0.021 & 3.8   &  4.0  \\
  42.8 &   8412 &27.1 &  0.023 & 3.5   &   3.9 \\
  46.1 &   9052 &29.1 &  0.024 & 3.3   &  3.5  \\ 
  49.4 &   9692 &31.2 &  0.026 & 3.0   &  3.3 \\ \hline
\end{tabular} 
\caption{Encounters with a single debris swarm  of mass $10^{17}$~g created at true anomaly $135^\circ$, 80 days after perihelion passage. The fragments have initial random speeds and directions relative to the comet nucleus, uniformly distributed with maximum initial dispersion speed of $\rm 2 \,m\,s^{-1}$. The trail length $L$ is in units of $R_\oplus$, the duration $D$ of crossing the Earth's orbit is in days, and the encounter probability is $\nu$. The cumulative probability of an encounter with this swarm is ${\sim}0.2$. The mass $M_e$ intercepted is in units of $10^{13}$~g, and the kinetic energy $E$ of the incident material is in units of $10^3$ megatons.}
\label{tab:enc}
\end{table}     

When the orbits of comet and Earth intersect, a fragment swarm of length $L$ will cross the Earth's orbit with duration of passage $D = L/V_o$ days, being a fraction $D/P$ of its orbital period.  The probability $\nu$ of a terrestrial encounter during such a passage is
\begin{equation}
\nu = L/(V_oP) {\sim}2.68\times 10^{-6} L
\end{equation}
\noindent where $\rm V_o{\sim}23~ km\,s^{-1}{\sim}310.5\,R_\oplus/day$. Table~4 illustrates the encounter probabilities per passage of a $10^{17}$\,g swarm of fragments with initial random dispersion speeds of up to $\rm 2 \,m\,s^{-1}$. The weighted mean length of the swarm over 50~yr is ${\sim}6680\,R_\oplus$ and the cumulative probability that the Earth will encounter it while $E>6000$\,Mt is $p{\sim}$0.2.  

The mean mass $M_e$ intercepted by the Earth during passage through this swarm was determined by following the evolution of 50,000 particles representing it, and counting the numbers intercepted by the Earth during each of 100 random passages through it. The encounter energy lay in the range 6000-50,000 megatons over a 50~yr period following the splitting. During this time another 8-16 or so splittings may take place ($f_0$ of Table~1), adding to the density of the expanding trail and hence the encounter energy. If nodal crossings occur 10 times in the course of a 15,000~yr disintegration history, consistently with the precession period of Comet Encke, then there is only a 10\% chance, $(1-p)^{10}$, of avoiding an encounter with mean impact energy $<E>\sim 6000$\,Mt.

The probability of a damaging encounter can also be roughly estimated by scaling up from the single-body collision probability (e.g. Kessler 1981, Rickman et al. 2014). Taking the interval between collisions with a body like Comet Encke as $5\times 10^8$~yr,
then the interval $\Delta t$ in years between encounters with a fragment swarm is of order
\begin{equation}
\Delta t = \frac{5\times 10^8}{A} \times \frac{l_c}{l_f} \times \frac{1}{N}
\end{equation}
\noindent where $A$ is the face-on area of the swarm in units of the cross-section of the Earth, $l_c$ is the active lifetime of the comet, $l_f$ is the effective lifetime of the swarm as a significant hazard, and $N$ is the number of such fragments generated over the active lifetime of the comet. We have $A = L \cos\chi\times 2\delta z$. Here $\delta z$ is either the out-of-plane dispersion of the fragments, determined by a least-squares fit to the computed trail, or the radius of the Earth, whichever is larger (generally the latter for modest dispersion speeds). Taking, from Table~4, a characteristic $L=6700\,R_\oplus$,  $l_c = 10^4$\,yr, $l_f = 50$\,yr and $N = 10^3$,  eqn (5) yields an interval  $\Delta t {\sim} 17,000$~yr between encounters of at least 6500 Mt energy, and 3,800~yr between encounters of at least 5000\,Mt energy.  There is thus good agreement between the encounter rates obtained from eqns (4) and (5). Orbital precession is significant over the millennia during which the comet is fragmenting, and second or third generation comets may themselves form individual branches, but this does not affect the overall encounter probabilities as estimated above. 

Both planetary perturbations and solar radiation pressure will act to disperse a meteor stream. The simulations here apply to fragments larger than the micron-sized dust particles associated with tails. Spitzer observations of the fragments associated with the breakup of Comet 73P/Schwassmann-Wachmann 3 appear to show that their mass is dominated by particles 0.1~mm size upwards (Vaubillon \& Reach 2010), which are less subject to Poynting-Robertson drag over the timescales considered here (Williams 2002). Comet Encke lies outside the Jovian sphere of influence, and Jupiter's differential influence on a fragment swarm over say a dozen revolutions may also be neglected for the present purpose.

\section{Multiple impacts and climate transitions}
The disintegration history of comets is varied and has been extensively discussed (e.g. Boehnhardt 2004; Fernandez 2009). Comets may split into a few fragments, or produce many small pieces, or disintegrate altogether, but cascading fragmentation is probably the prime disintegration process, and the existence of many meteoroid substreams within the Taurid complex indicates that the progenitor comet took this route. Unseen remnants (dormant comets) may be a major source of meteor showers (Jenniskens 2008). At least 19 of the brightest high-eccentricity near-Earth asteroids have been statistically linked to the Taurid complex (Napier 2010), along with others (Spurny et al. 2017; Porub\u{c}an et al. 2004). Swarm material will be in the form of dust, boulders and probably larger bodies, and the duration of encounters will range from a few hours to a day or two. As the cluster decays from a handful of bodies to a meteor stream, intermediate stages are likely to involve increasing numbers of smaller bolides. The spectra and breakup heights of Taurid meteors show a variety of compositions and strengths, consistent with a heterogeneous composition of the primary body (Matlovi\u{c} et al. 2017; Tubiana et al. 2015), and it is likely that larger fragments will yield Tunguska-like airbursts. If we consider that a wildfire could be initiated by a one megaton bolide impacting on suitable material, and half the mass of say a 10000 megaton encounter (Table~\ref{tab:enc}) was in the form of such bolides, then a global wildfire would likely be initiated.

Studies of Earth history of the last 150,000 years have revealed that the climate is subject to sudden temperature changes, with transitions often taking place within decades or even a few years (Flohn 1979; Adams et al 1999; Steffensen et al 2008), and persisting for a few centuries up to a millennium or so. These changes are superimposed on the longer climatic cycles probably caused by orbital and polar precessions (Rial 1999). The onset of the Younger Dryas cooling was one such event. It was abrupt in onset and intense, amounting to 2 to 6$^\circ$ in the Northern hemisphere, and was accompanied by a sudden transition in European vegetation from temperate to Scandinavian. In modern times, a sudden global cooling of this amplitude would have a calamitous effect on agriculture (Engvild 2003). 

The abrupt collapse of the Akkadian and other civilisations around 4170 BP has been attributed to a shift to more arid conditions, drought driving subsequent societal and political collapse (Cullen et al. 2000). The onset of the aridity has been attributed to a cooling of the North Atlantic. Several hypotheses have been proposed to explain such abrupt coolings and there are likely to be multiple causes given the complexity of internal mechanisms. Courty et al. (2007), on the basis of paleosoil evidence, have proposed that the Akkadian collapse was due to a cosmic disturbance (cf Coqueueniot \& Courty 2012). We discuss here the possibility that the sudden and extreme coolings recorded over the Quaternary may have arisen from comet dust sprinkling the upper atmosphere and reducing sunlight incident on the Earth's surface (Clube \& Napier 1984; Asher et al. 1994; Clube et al. 1996; Napier 2001).  

The hypothesis that cosmic dusting might be responsible for ice ages has a long history. Passage through molecular clouds has been proposed as the cause of ice epochs or snowball glaciations (McCrea 1975, 1981; Begelman \& Rees 1977; Yabushita \& Allen 1985, 1989; Pavlov et al 2005; Kataoka et al. 2014). Hoyle \& Wickramasinghe (1978) raised the possibility that ecological catastrophes such as ice ages and the Cretaceous-Tertiary extinctions may have been caused by passages close to the nucleus of a comet. The proposed timescales for these events are geological. However, with the finding that large Centaurs leak into the near-Earth environment at a geologically rapid rate (Galiazzo et al. 2019, Napier 2015, Horner et al. 2014), it now appears that such bodies are likely to be suppliers of cosmic dust on much shorter timescales, extending into the Quaternary and even historical times. 

A 100~km comet with density $\rm 0.4~g\,cm^{-3}$, comprising 50\% dust, releases $10^{14}$~tons of dust over the course of its disintegration, largely through the intermediary of hierarchic fragmentation. The debris encountered during these passages will generally be a mixture of dust and larger fragments. From Table~4 and the associated discussion it appears that the Earth may intercept $10^{7}$~tons of dust and boulders over a few hours during such an encounter, as against 30-50 tons of normal daily background flux. Much of this meteoric input will ablate to smoke in the mesophere, i.e. micron-sized particles (Klekociuk et al. 2005). Particles with mass $\ltsimeq 10^{-12}$g will not become hot enough to ablate. An organic grain of diameter $1 \mu$ and density $2 \rm g\,cm^{-3}$ has a mass $10^{-14}$~g yielding, for a uniform distribution of dust over the Earth's atmosphere, a square centimetre column of $2\times 10^8$ particles. For spherical particles, the intensity of sunlight reaching the ground is then reduced by a factor $\exp (-\tau)$ with  optical depth $\tau{\sim}1.6Q_{ext}$ for vertical incidence of sunlight. $Q_{ext}\sim$2 for particles $\gtsimeq 1~\mu$ in diameter leading to a diminution of sunlight at ground level by factor of $\sim$25.  The carbon content of cometary dust has been measured in Comets 1P/Halley and 67P/Churyumov-Gerasimenko, and has been found to be high (e.g. Jessberger et al. 1988 for 1P and Bardyn et al. 2017 for 67P). Thus the dust particles of Comet 67P are made of 50 percent organic carbon by mass, presumably reflecting the composition of pristine solar system material. Organic carbon spheres have a high scattering efficiency, ${\sim}1.5$, and low absorption efficiency, ${\sim}0.05$, diminishing sunlight by a factor of 10. The creation of such meteoric smoke in the mesosphere at the concentrations considered here would effectively turn the Earth white in visible light, while allowing infrared radiation to escape from the surface (Hoyle \& Wickramasinghe 1978), during the months or years of settling. Coagulation of aerosols would, however, change the shape and size of the particles, and introduce porosity, all of which would alter their absorption and scattering properties. The extent to which charged aerosols would coalesce is uncertain: in noctilucent cloud conditions, in situ rocket observations have revealed both larger and smaller particle radii (L\"{u}bken \& Rapp 1981). The settling time of spherical micron-sized particles is about 3 to 5 years, depending on factors such as latitude and season. The overall climatic effects are likely to be complex (Renssen et al 2015) and to involve an altered atmospheric circulation, and the blocking of sunlight by the soot generated from the intense wildfire activity at the YDB, which yields an essentially infinite optical depth during its residence time of days to weeks in the lower atmosphere (Wolbach et al. 2018). 

The injection of ${\sim}10$ million tons of carbon-rich aerosols and water vapour into the mesosphere over a few hours (Table 4), as against the current background flux of ${\sim}50$ tons/day, is likely to lead to noctilucent clouds, an increased planetary albedo, and significant cooling. One would expect the climate (but not necessarily the biosphere) to recover within a decade (Robock et al. 2009) unless the cooling triggers a climate instability such as that proposed by Hoyle (1982). 

\section{Conclusions}
I have modelled the disintegration of a large comet in a short-period orbit, using Comet Encke as an archetype, and find that there is a reasonable expectation of one or more brief meteor `hurricanes', with intensities far beyond modern experience, in the course of disintegration of the progenitor. Enough meteoric smoke may be created during such encounters to generate sudden coolings of some years' duration, along with widespread wildfires. The terrestrial upsets at the onset of the Younger Dryas boundary of 12,900 BP, and  the simultaneous collapse of early civilisations around 2350 BC, may have been triggered by events of this character.

\section{Acknowledgements}
I thank David Asher for discussions on this topic.

\end{document}